# Spin relaxation in a nanowire organic spin valve: Observation of extremely long spin relaxation times

S. Pramanik, C-G Stefanita, S. Patibandla and S. Bandyopadhyay[1]

Department of Electrical and Computer Engineering

Virginia Commonwealth University, Richmond, Virginia 23284, USA

K. Garre, N. Harth and M. Cahay

Department of Electrical and Computer Engineering and Computer Science

University of Cincinnati, Cincinnati, Ohio 45221, USA

ABSTRACT

We report spin valve behavior in an organic nanowire consisting of three layers - cobalt, $Alq_3$ and nickel – all nominally 50 nm in diameter. Based on the data, we conclude that the dominant spin relaxation mechanism in $Alq_3$ is the Elliott-Yafet mode. Despite the very short momentum relaxation time, the spin relaxation time is found to be very long – at least a few milliseconds - and relatively temperature independent up to 100 K. To our knowledge, this is the first demonstration of an organic nanoscale spin valve, as well as the first determination of the primary spin relaxation mechanism in organics. The unusually long spin relaxation time makes these materials ideal platforms for some areas of spintronics.

---

[1] Communicating author. E-mail: sbandy@vcu.edu

π-conjugated organic semiconductors are an important platform for 'spintronics' that purports to harness the spin degree of freedom of a charge carrier to store, process, and/or communicate information[1]. Spin orbit interaction in organics is typically very weak, which should result in long spin relaxation times[2]. Many organics are also optically active[3] and therefore could lead to multi-functional "opto-spintronic chips" where optics and spintronics are integrated to perform seamless signal processing and communication functions. Such chips will be inexpensive, versatile, and the tremendous flexibility afforded by synthetic organic chemistry offers limitless possibilities in terms of the variety and complexity of structures that can be realized. Already some efforts have been made to combine optics with spintronics in organics[4].

Recently, a thin-film organic spin valve structure consisting of an organic semiconductor placed between two ferromagnetic electrodes was demonstrated[1]. Some theoretical effort has also been made to understand spin transport in such organics[5, 6], but any insight into the primary spin relaxation mechanism is still lacking.

There are four major spin relaxation mechanisms in semiconductors: the D'yakonov-Perel' (D-P)[7], the Elliott-Yafet (E-Y)[8], hyperfine interaction between nuclear and carrier (electron or hole) spins[9], and the Bir-Aronov-Pikus (B-A-P) mechanism[10]. They dominate in both semiconductors and metals[11].

It is important to establish which of these four mechanisms is the most dominant in organics. The two likely candidates are the D-P and the E-Y mechanisms since the B-A-P mechanism is absent in unipolar transport and the hyperfine interaction is very weak in organics. The D-P mechanism is suppressed by quasi one-dimensional confinement[12]. Therefore, if the relaxation rate is found to decrease upon confining carriers to a quasi one dimensional structure, then we will have established that the primary mechanism is the D-P mode. On the other hand, the E-Y mechanism can be exacerbated by quasi one-dimensional confinement if the latter increases the momentum relaxation rate. Thus, any increase in the spin relaxation rate upon quasi one dimensional confinement is a strong indicator that the E-Y mechanism is dominant.

Based on this premise, we have fabricated a nanowire spin valve structure consisting of three layers – cobalt, Alq$_3$ [tris-(8-hydroxy-quinolinolato) aluminum] and nickel. The structures were synthesized by using a porous alumina membrane containing a well ordered hexagonal close packed arrangement of pores with 50 nm diameter. The fabrication of such films has been described in ref. 13. It is produced on an aluminum foil. There is an alumina "barrier layer" at the bottom of the pores which is removed by a reverse polarity etching technique[14]. Nickel is then electrodeposited selectively within the pores from a solution of $NiSO_4:6H_2O$ using a dc voltage of 1.5 V. Next, Alq$_3$ is evaporated on the porous film through a mask with a window of area 1 mm$^2$ in a vacuum of $10^{-7}$ Torr. The Alq$_3$ seeps into the pores and reaches the nickel. The fact that Alq$_3$ is a short stranded organic of low molecular weight is helpful in transporting it inside the pores. The thickness of the evaporated Alq$_3$ layer is monitored by a crystal oscillator. Finally, cobalt is evaporated on the top without breaking vacuum (as in Ref. 1). The resulting structure is schematically depicted in Fig. 1.

For electrical measurements, we attach two gold wires to the top cobalt and the bottom aluminum layers. Since the contact area on top is 1 mm$^2$ and the nanowire density is $2 \times 10^{10}$ cm$^{-2}$, the contact pads cover about $2 \times 10^8$ wires in parallel. Of course not every wire makes electrical contact so that the actual number of wires contributing to the measured resistance is much smaller. Magnetoresistance of the structure is measured in a Quantum Design Physical Property Measurement System. Typical magnetoresistance traces at three different temperatures are shown in Fig. 2 where the magnetic field is parallel to the axis of the wires. This direction also corresponds to the easy axis of magnetization for the nickel and cobalt nanomagnets within the pores. There is a background *monotonic* magnetoresistance which is often observed in these structures because of the anisotropic magnetoresistance effect[15], but more importantly, we find magnetoresistance peaks located between fields of 800 Oe and 1800 Oe which are the coercive fields of the nickel and cobalt nanowires. This is the tell-tale signature of the spin valve effect. The height of this peak decreases with increasing temperature and is barely visible at 100 K.

In these structures, it is not possible to measure the coercivities of the cobalt and nickel contacts individually using conventional techniques. SQUID measurements do not resolve the coercivities. However, we had individually measured the coercivities of nickel and cobalt nanowires in the past[13, 16]. For nickel, ref. 13 reported a maximum coercivity of 950 Oe at room temperature for nanowires of diameter 18 nm and it decreased to 600 Oe for wider nanowires of 21 nm diameter. Since coercivity increases with decreasing temperature[17], a value of 800 Oe is quite possible in 50-nm diameter nanowires at 1.9 K. The coercivity of cobalt nanowires has been studied extensively in ref. 16. The coercivity of 22 nm diameter wires was found to be > 1600 Oe at room temperature, so that the coercivity of 50 nm wires can quite likely be 1800 Oe at the low temperature of 1.9 K. Thus, the leading and trailing edges of the peaks in Fig. 2 seem to occur at the coercive fields of the ferromagnetic contacts.

From the relative height of the resistance peak $\Delta R/R$ shown in Fig. 2, we can extract the spin diffusion length in the Alq3 layer following the technique employed in ref. 1. We first assume that there is no loss of spin polarization at the interface between Alq3 and the injecting ferromagnetic contact because of the so-called self adjusting capability of the organic[1, 18]. Next, we assume, as in ref. [1], that there is a potential barrier at the organic/ferromagnet interface that the injected carriers tunnel through with a surviving spin polarization $P_1$. This barrier could be the Schottky barrier due to the contact potential. After this, the carriers drift and diffuse through the remainder of the organic layer under the influence of the electric field, with exponentially decaying spin polarization $exp[-(d-d_0)/\lambda_T]$ where $d$ is the total width of the organic layer, $d_0$ is the spatial extent of the potential barrier, and $\lambda_T$ is the spin diffusion length in Alq3 at a temperature $T$. The Schottky barrier at the detecting contact is lowered by the electric field and therefore does not present a potential barrier for tunneling. This picture is adapted from ref. [1]. Finally, if the spin polarization at the Fermi level of the detecting contact is $P_2$, then $\Delta R/R$ is given by the Julliere formula[19]

$$\frac{\Delta R}{R} = \frac{2 P_1 P_2 e^{-(d-d_0)/\lambda_T}}{1 - P_1 P_2 e^{-(d-d_0)/\lambda_T}} \qquad (1)$$

We will now assume that $d_0 << d$. Later we will show that this assumption is valid. In that case, the loss of spin polarization in tunneling through the potential barrier is negligible. Therefore, $P_1$ is approximately the spin polarization of the injecting contact. Since the spin polarization in cobalt and nickel at their Fermi energies are 42% and 33% respectively[20], $P_1 = 0.42$ and $P_2 = 0.33$.

In order to determine the value of *d*, we have carried out transmission electron microscopy (TEM) of the nanowires. The wires were released from their alumina host by dissolution in very dilute chromic/phosphoric acid, washed, and captured on TEM grids for imaging. The TEM micrograph of a typical wire is shown in Fig. 3 (a). The $Alq_3$ layer thickness *d* is found to be 33 nm, which is quite close to the layer thickness estimated from the crystal oscillator used to monitor thickness during the evaporation of $Alq_3$ (that value was 30 nm). This agreement gives us confidence that *d* does not vary too much from one wire to another. We assume that it varies by ± 5 nm when we calculate $\lambda_T$. This will introduce some uncertainty in the spin diffusion length.

Current voltage characteristics of the nanowires are shown in the inset of Fig. 3. They are symmetric because of equal coupling to the contacts[21], but non-linear between -3.5 and 3.5 V at all measurement temperatures, indicating that the contacts are Schottky in nature. This means there has not been significant inter-diffusion of Co or Ni into the $Alq_3$ layer, since that would have produced an ohmic contact. As a result, the layer thickness *d* is well defined in the nanowires, which allows us to apply Equation (1) to estimate $\lambda_T$.

In estimating $\lambda_T$ from Equation (1), we assumed that $d_0 - d \cong d$. If this approximation is valid, then the estimated $\lambda_T$ will be independent of *d*. To confirm that fact, we fabricated another set of samples with slightly smaller *d*. Fig. 3(b) shows the TEM micrograph of a wire from this set where the layer thickness is found to be 26 nm. The quantity $\lambda_T$ measured from this set at any temperature agrees to within ~10% with that measured from the other set at the same temperature. Therefore, $\lambda_T$ is reasonably independent of $d$. The values of $\lambda_T$ as a function of temperature are plotted in Fig. 4(a).

Comparing the measured values of $\lambda_T$ to those reported in thin films of Alq$_3$ (45 nm at 4.2 K)[1], we find that quasi one dimensional confinement has actually *reduced* $\lambda_T$ by almost an order of magnitude. If the D-P mechanism were the primary cause of spin relaxation, then $\lambda_T$ should have increased. Since we find the opposite trend, we conclude that the primary relaxation mechanism is the E-Y mechanism.

Elliott has derived a relation between the spin relaxation time $\tau_s$ and the momentum relaxation time $\tau_m$ [11] which Yafet has shown to be temperature independent [22]:

$$\frac{\tau_m}{\tau_s} \propto \frac{\Delta}{E_g} \qquad (2)$$

Here $\Delta$ is the spin orbit interaction strength in the band where the carrier resides (in our case the LUMO band) and $E_g$ is the energy gap to the nearest band (in our case the HOMO-LUMO gap). Since $\tau_s(T) = \lambda_T^2/D(T) = m^*\lambda_T^2/(kT\tau_m)$, where $D(T)$ is the temperature dependent diffusion coefficient related to the mobility by the Einstein relation and $m^*$ is the effective mass, Equation (2) can be recast as $(\tau_m^2 kT/m^*\lambda_T^2) \propto \Delta/E_g$. Since neither $\Delta$ nor $E_g$ is affected by quasi one-dimensional confinement, we can posit that at any temperature

$$\frac{\tau_m^{2D}}{\tau_m^{1D}} = \frac{\lambda_T^{2D}}{\lambda_T^{1D}} \qquad (3)$$

where '2D' refers to thin film, and '1D' refers to nanowire. From Equation (3), we find that one-dimensional confinement has reduced $\tau_m$ by a factor of ~ 10. This is possible in our structures.

There is a huge density of charged surface states - of the order of $10^{13}$/cm$^2$ - at the interface between the nanowire and its ceramic host (alumina)[23]. These surface states will cause frequent momentum randomizing collisions in the nanowire via Coulomb interaction, which will significantly reduce $\tau_m$ in nanowires compared to thin films.

It is possible to estimate the temperature dependent spin relaxation *time* $\tau_s(T)$ from $\lambda_T$ using the relation $\tau_s(T) = \lambda_T^{\ 2}/D(T) = e\lambda_T^{\ 2}/kT\mu$, where $\mu$ is the drift mobility. The reported drift mobility in Alq$_3$ is given by the relation[24]:

$$\mu(E) = \mu_0 \exp[\alpha E^{1/2}] \quad (4)$$

where $\mu_0$ and $\alpha$ are constants and $E$ is the electric field. Ref. 24 reports $\mu_0 = 10^{-7} – 10^{-9}$ cm$^2$/V-sec, and $\alpha = 10^{-2}$ (cm/V)$^{1/2}$ in the *bulk* organic.

In order to determine the electric field $E$ in the organic, we proceed as follows. The voltage over the nanowires can be estimated from the measured resistance and the current using Ohm's law: V = IR = 10 µA x 1520 Ω = 15.2 mV. Since the Alq$_3$ layer (in the first set) is nominally 33 nm wide, the average electric field across it is 15.2 mV/33 nm = 4.6 kV/cm. Using this value in Equation (4), we estimate that the carrier mobility in the *bulk* organic is 2 x 10$^{-7}$ – 2 x 10$^{-9}$ cm$^2$/V-sec. In nanowires, the mobility is 10 times lower since we found that the momentum relaxation time is 10 times smaller. Therefore, our mobility is 2 x 10$^{-8}$ – 2 x 10$^{-10}$ cm$^2$/V-sec.

Assuming that the mobility is temperature independent, we have calculated the spin relaxation time $\tau_s(T)$ from the relation $\tau_s(T) = e\lambda_T^{\ 2}/kT\mu$. These results are plotted as a function of temperature in Fig. 4(b). The two curves give the maximum and minimum values of $\tau_s(T)$ at different temperatures. They range from few milliseconds to over 1 second at 1.9 K. These are among the longest spin relaxation times reported in any system.

In conclusion, we have demonstrated the first 'quantum wire' organic spin valve, and in the process identified the dominant spin relaxation mechanism in organics to be the E-Y mode. We have also demonstrated that the spin relaxation time in organics is *exceptionally long* which is consistent with vanishingly small spin orbit interaction strength in organics. This establishes organic semiconductors as a very viable platform for spintronics.

**Acknowledgement**: The work at Virginia Commonwealth University is supported by the US Air Force Office of Scientific Research under grant FA9550-04-1-0261.

**Figure captions:**

Fig. 1: Schematic representation of a nanowire spin valve structure.

Fig. 2: Magnetoresistance traces of the nanowires with a ~ 33 nm $Alq_3$ layer. The magnetic field is parallel to the axis of the wires: (a) at a temperature of 100 K, and (b) at temperatures of 1.9 K and 50 K. Solid and broken arrows indicate reverse and forward scans of the magnetic field. The parallel and anti-parallel configurations are shown in Fig. 2(b).

Fig. 3: Transmission electron micrograph of a typical nanowire spin valve structure from (a) the first set showing that the $Alq_3$ layer width is about 33 nm, and (b) the second set showing that the $Alq_3$ layer width is about 26 nm. The inset shows the current-versus voltage characteristic at three different temperatures.

Fig. 4: (a) The spin diffusion length, and (b) the spin relaxation time, in nanowire $Alq_3$ as a function of temperature. The triangles are data obtained from the first set and the squares are data obtained from the second set. We show the maximum and minimum values at different temperatures. In the case of the spin diffusion length, the non-zero range comes from the ± 5 nm uncertainty in the organic layer thickness, while in the case of the spin relaxation time, the range accrues mostly from the large uncertainty in the mobility.

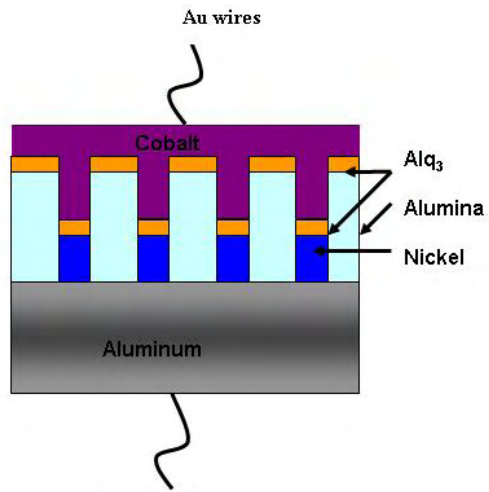

Fig.1.

Pramanik, et al.

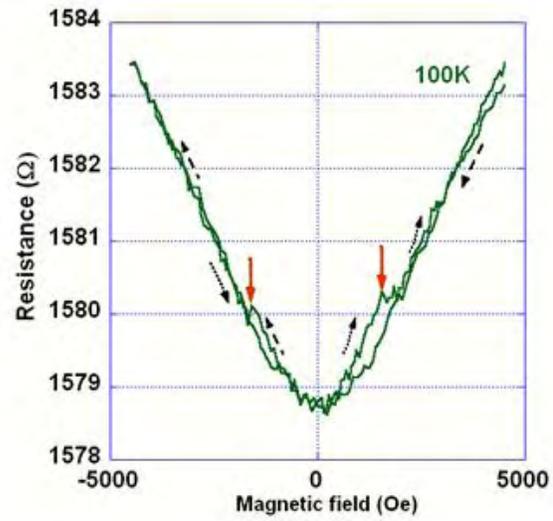

(a)

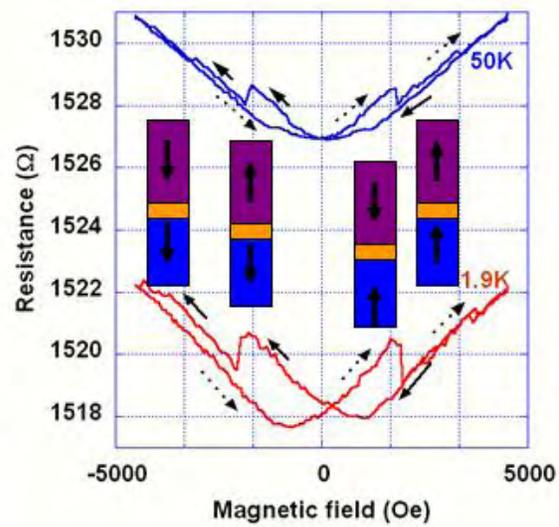

(b)

Fig. 2. Pramanik, et al

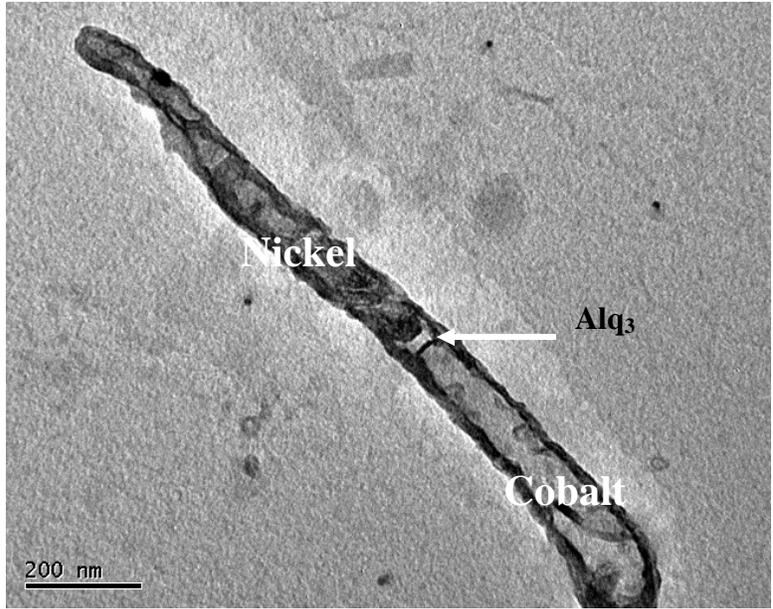

(a)

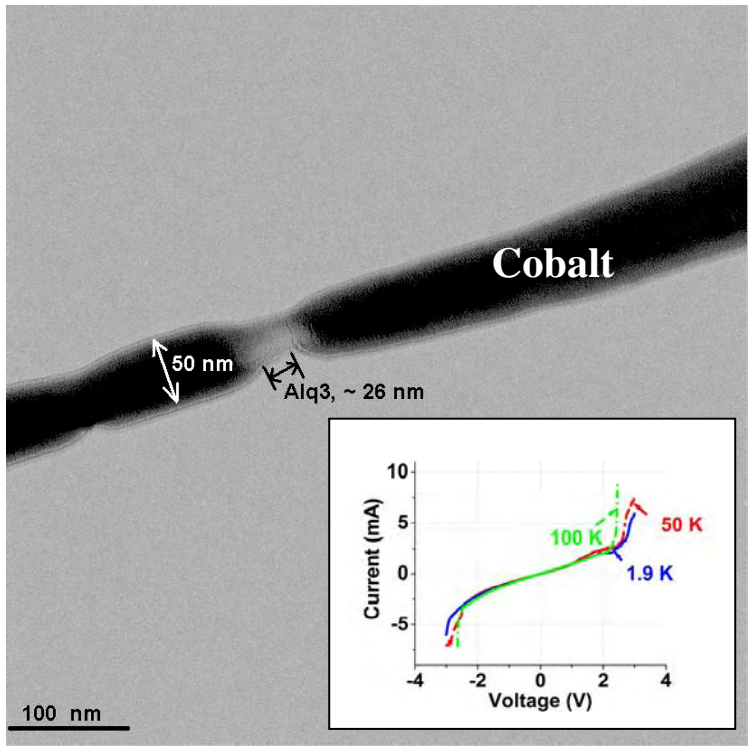

(b)

Fig. 3: Pramanik et al

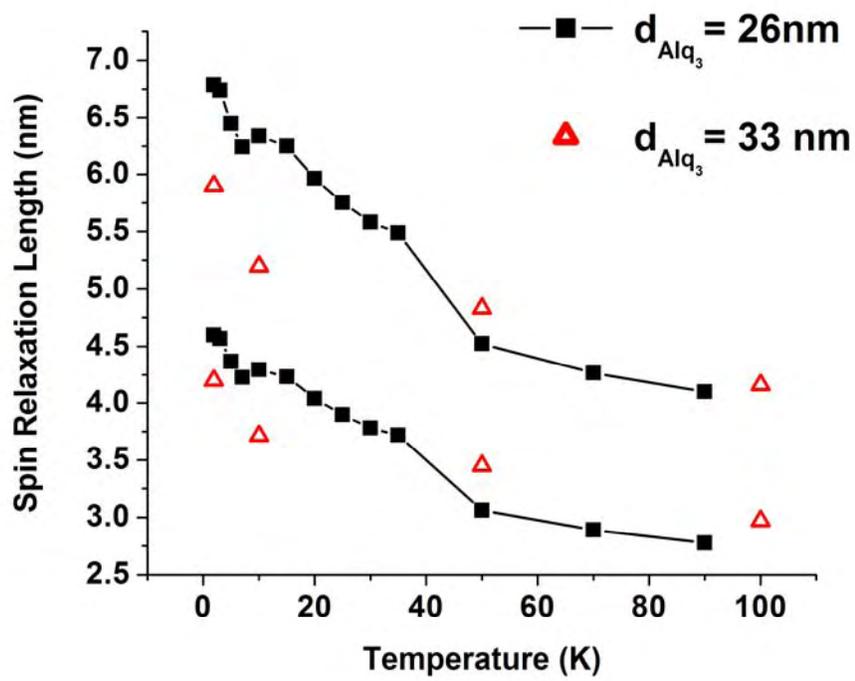

(a)

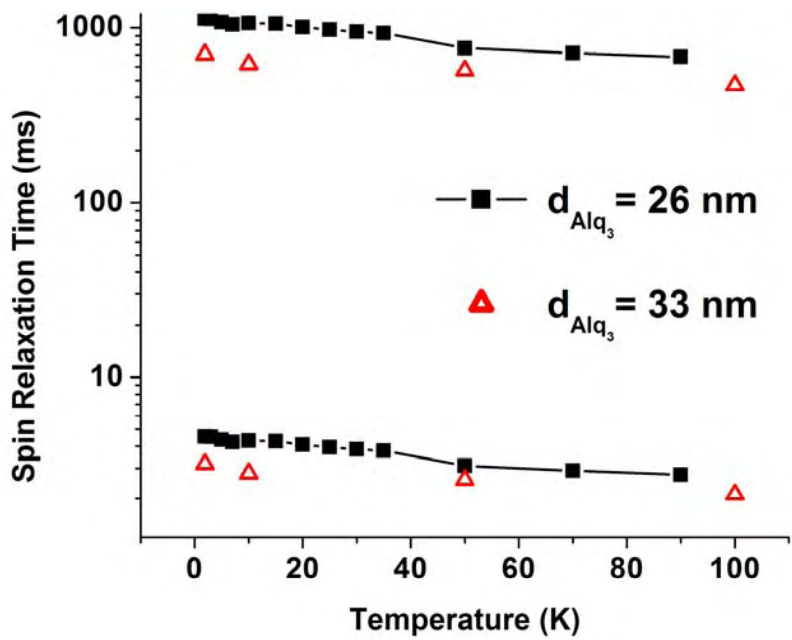

(b)

Fig. 4. Pramanik et al.